\documentclass[conference]{IEEEtran}
\usepackage{amssymb,amsmath,amsthm}
\usepackage{caption}
\usepackage{cancel}
\usepackage{mdwmath}
\usepackage{mdwtab}
\usepackage{soul}
\usepackage{cite}
\usepackage{multirow}
\usepackage{color} 
\usepackage{setspace}
\usepackage{esvect}
\usepackage{amsmath}
\usepackage{algorithm,algorithmic, multicol}

\usepackage{subfig}

\ifCLASSINFOpdf
  \usepackage[pdftex]{graphicx}
  \graphicspath{{./}}
  \DeclareGraphicsExtensions{.pdf,.jpg,.png,.xps}
\else
\fi

\begin{document}

\title{Replication of Virtual Network Functions: Optimizing Link Utilization and Resource Costs}

\author{\IEEEauthorblockN{Francisco Carpio, Wolgang Bziuk and Admela Jukan}
	\IEEEauthorblockA{Technische Universit{\"a}t Braunschweig, Germany}
	\IEEEauthorblockA{Email:\{f.carpio, w.bziuk, a.jukan\}@tu-bs.de}
}

\maketitle

\begin{abstract}
Network Function Virtualization (NFV) is enabling the softwarization of traditional network services, commonly deployed in dedicated hardware, into generic hardware in form of Virtual Network Functions (VNFs), which can be located flexibly in the network. However, network load balancing can be critical for an ordered sequence of VNFs, also known as Service Function Chains (SFCs), a common cloud and network service approach today. The placement of these chained functions increases the ping-pong traffic between VNFs, directly affecting to the efficiency of bandwidth utilization. The optimization of the placement of these VNFs is a challenge as also other factors need to be considered, such as the resource utilization. To address this issue, we study the problem of VNF placement with replications, and especially the potential of VNFs replications to help load balance the network, while the server utilization is minimized. In this paper we present a Linear Programming (LP) model for the optimum placement of functions finding a trade-off between the minimization of two objectives: the link utilization and CPU resource usage. The results show how the model load balance the utilization of all links in the network using minimum resources.

\end{abstract}

\IEEEpeerreviewmaketitle

\section{Introduction}

Network Function Virtualization (NFV) is a new paradigm that virtualizes the traditional network functions and places them into generic hardware and clouds, as opposed to the designated hardware. The placement of the virtual network functions (VNFs) can happen either in remote data centers or by deploying single servers or clusters of servers. Placing VNFs in remote data center can lower the cost of deployment, but is known to typically increasing the delay and create churns of network load, due to the fix and often remote location. Installing new services (or, mini data centers) inside the network can mitigate the distance-to-datacenter problem. At the same time, the deployment of new servers forming small data centers in regular nodes requires new investment costs, which requires a gradual upgrade of the network. Therefore, the optimal placement of these servers in the network is a must for network operators to reduce the operational costs.

While most of the current work concentrates on the optimal placement of VNFs under some specific objective minimizing the network costs under some specific resources’ constraints, e.g. costs of power consumption or the number of physical servers, less effort has been on addressing the network load balancing problem with VNF placement. In this paper, we address, jointly, the network load balancing and resource cost problem with VNF placement, where the concept of VNF replications is used to find a trade-off between network  load balancing and network costs. The Fig. 1 illustrates the idea, whereby we assume a service chain which is composed by non-replicable VNFs and, depending on the number of replicas, may be splitted to one or more  parallel sets of an ordered sequence of VNFs towards the service end-point. The VNF replicas in a service chain can be implemented as many time as needed in the network. The major advantage is to split the traffic flows in a controlled way such that network traffic load balancing can be optimized when the service is running. As commonly in network services, the service chain starts with a non-replicable VNF, e.g. a load balancer or gateway, and is allocated in a dedicated data center which generates service requests. While the rest of the functions are allocated on small servers, maintaining the sequence order. To find the optimum placement of servers required for the deployment of VNFs, we formulate the problem as a Integer Linear Programming (ILP) model.

The rest of the paper is organized as follows. Section II presents related work. Section III describes the reference architecture. In Section IV, the related optimization model is described. Section VI analyzes the performance, and Section VII concludes the paper and discusses future research.

\section{Related Work}

Early work in \cite{Moens2015} studies the optimal VNFs placement in hybrid scenarios, where some network functions are provided by dedicated physical hardware and some are virtualized, depending on demand. They propose an ILP model model with the objective to minimize the number of physical nodes used, which limits the network size that can be studied due to complexity of the ILP model. In \cite{Mehraghdam2014}, a context-free language is proposed for the specification of VNFs and a Mixed Integer Quadratically Constrained Program (MIQCP) for the chaining and placement of VNFs in the network. The paper finds that the VNF placement depends on the objective, such as latency, number of allocated nodes, and link utilizations. In mobile core networks, \cite{Basta2014} discuss the virtualization of mobile gateways, i.e.,  Serving Gateways (S-GWs) and Packet Data Network Gateways (P-GWs) hosted in data centers. They analyze the optimum placements by taking into consideration the delay and network load. In \cite{Bagaa2014} also propose the instantiation and placement of PDN-GWs in form of VNFs. 
 
Unlike previous work, we solve the VNF placement problem by considering  VNF replications which is the novel idea that we already proposed in \cite{Carpio2016a}. In this work, we extend \cite{Carpio2016a} to solve the optimum placement of VNFs by minimizing the network cost while maximizing load balancing with constraints on the available resources in mobile core networks. Therefore, we propose a more realistic approach to enable an scalable growth of the mobile data traffic over years. This paper also includes a new traffic model which also considers the end-user traffic generated in the Radio Access Network which previous paper did not consider. We also consider multiple VNFs placement per node with replications, which is new.

\section{Reference Architecture}

The NFV architecture is basically described by three components: Services, NFV Infrastructure (NFVI) and NFV Management and Orchestration (NFV-MANO). A \emph{Service} is the composition of VNFs that can be implemented in virtual machines running on operating systems or on the hardware directly. The hardware and software resources are provided by the NFVI that includes connectivity, computing, storage, etc. Finally, NFV-MANO is composed by the orchestrator, VNF managers and Virtualized Infrastructure Managers responsible for the management tasks applied to VNFs. 

In NFV-MANO, the orchestrator performs the resource allocation based on the conditions to perform the assignment of VNFs chains on the physical resources. The sub-task running in the orchestrator, known as VNF Forwarding Graph Embedding (VNF-FGE) or VNF placement problem, tries to find the optimum place to allocate VNFs with regard to some specific objective, such as minimization of computation resources, minimization of power consumption, network load balancing, etc. 

\begin{figure}[!t]
	\includegraphics[width=3.2in]{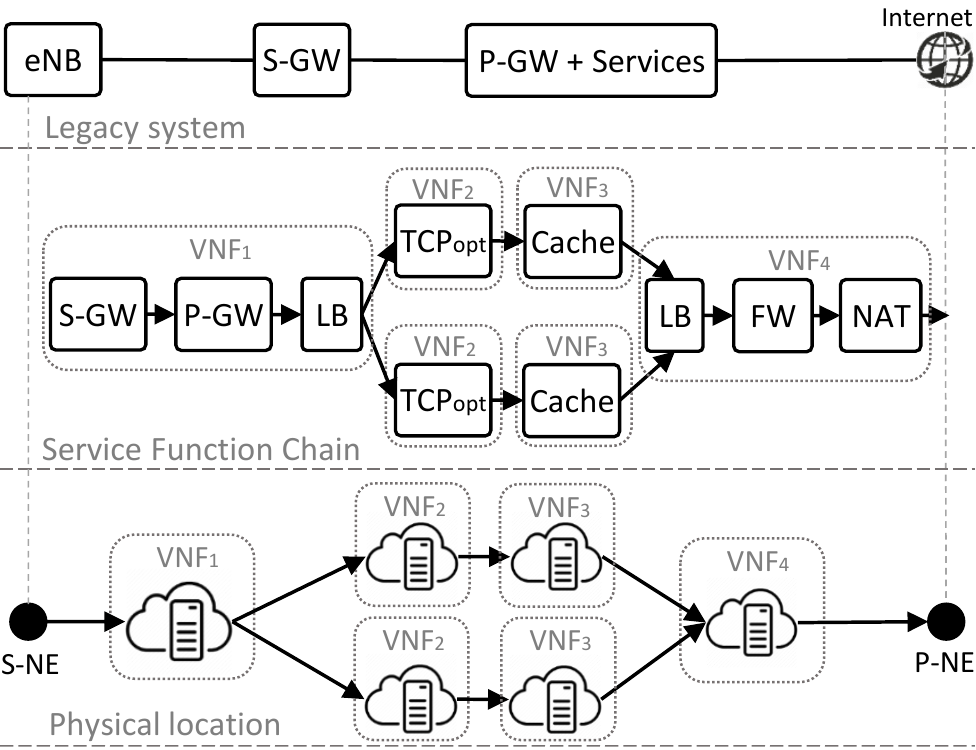}
	\caption{Use case on Mobile Core Networks}
	\label{chain}
	\vspace{-0.3cm}
\end{figure}

\subsection{Gateway Virtualization: The EPC Use-Case}


In the focus area of VNF placement problem, we propose to study a concrete scenario from legacy mobile systems, namely the Evolved Packet Core (EPC) networks, where we believe that the VNF placement in Service Function Chaining can bring most benefits. The proposed case study is shown in Fig. \ref{chain}. The Serving Gateway (S-GW) and PDN Gateway (P-GW) are connected to e-NodeB and send the end-user traffic towards Internet. This traffic usually requires various additional services, currently deployed using traditionally embedded network functions, such as load balancers, TCP optimizers, firewalls and NATs. Considering the virtualization of the S-GW and P-GW  on small data centers, as proposed in \cite{Basta2014}, both the data-plane and control-plane functions of the current legacy gateways are moved to an operator's datacenter. At the places of the legacy gateways, an off-the-shelf network element (NE) is used to redirect the traffic from the origin access point of the Radio Access Network (formely done by the S-GW) to the DC and from the DC to the external backbone interface, which is, in the most cases, the former place of the P-GW. Typical VNFs related to control-plane functionalities of the S-GW and P-GW can not be further distributed on the network. In contrast, due to the large traffic volumes handled by the data-plane, parallel transmission paths can be used in the transport network, and thus, VNFs related to data-plane functions with high intensives tasks, e.g. TCP optimizers (TCP\_opt) or Performance Enhancement Proxies (PEP),  may also be replicated and be used in parallel at different network locations, as proposed in \cite{draft-ietf-sfc-use-case-mobility-07}. Thus, we study the chaining and virtualization of the additional functions related to the data-plane on different physical locations (small data centers) in the mobile core network. 
The number of required replicas will be in relation with the network traffic demands. Therefore, by knowing how many replicas are necessary we can place them to maintain a good network load balancing. On the other hand, the usage of additional network locations can increase the number of required servers and DCs, which potentially will increase the network costs. To this end, we define the problem as finding the optimum placement for these functions subject to the network costs in form of used DC locations while at the same time load balancing the network. 

\begin{figure}[!t]
	\includegraphics[width=3.5in]{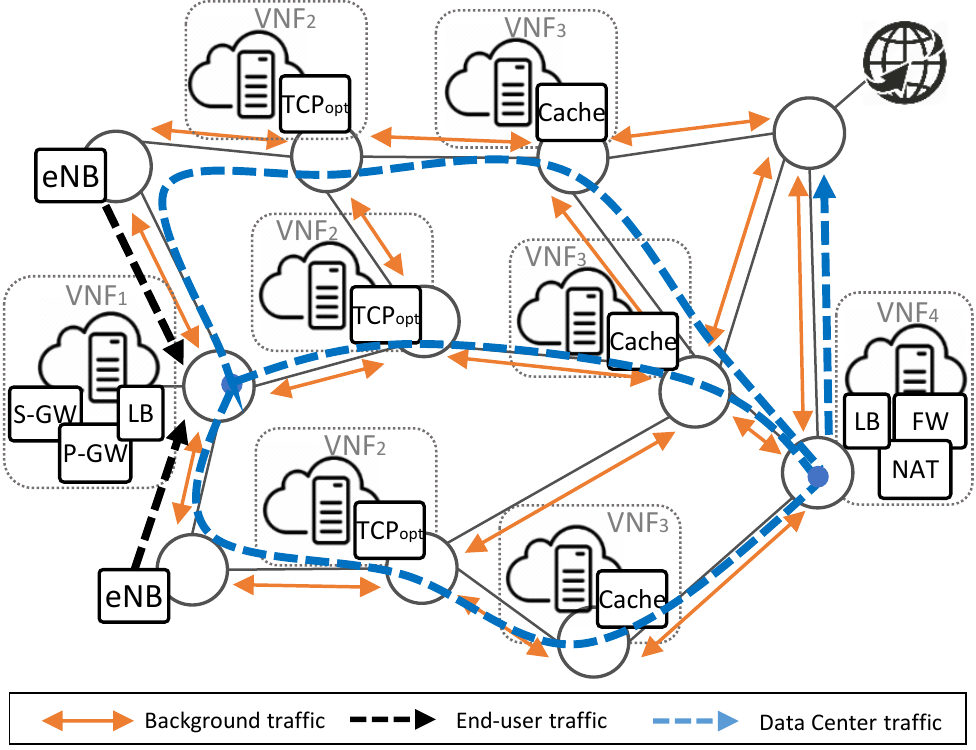}
	\caption{Network traffic considerations}
	\label{use-case}
	\vspace{-0.3cm}
\end{figure}

\subsection{Network Traffic Model}

In a real network, not all services will be virtualized at once. Thus in this paper, we assume two kinds of network traffic, defined as the \emph{background traffic} and \emph{data center traffic}. This is illustrated in Fig. \ref{use-case}. The background traffic is related to the legacy not virtualized services and the traffic is generated from each core node to the rest of nodes and routed by a traditional network core protocol, e.g. IP routing. The virtualized services are responsible for the data center traffic.  This traffic is generated at the S-NEs (former S-GW places) and has to be transported towards the Internet gateways (P-NE's at former P-GW locations), as shown in Fig. \ref{chain}. Since the latter category of traffic, usually TCP connections, is generated by end users, it has to traverse a set of network functions to match the required service before accessing to Internet.

In our approach, we assume that the background traffic can be generated randomly and forwarded following the rules of a specific Traffic Engineering (TE) model defined by the traditional single path destination oriented  IP routing. The TE model written in form of ILP formulation (detailed in the next section) minimizes the link utilization of all links in the network using a linear cost functions approach. Once the background traffic is load balanced, it will not be affected by the control of the data center traffic, but it has to be considered as a fixed input parameter for the next model called Resource Allocation (RA). This model is used to allocate optimally VNFs in the network trying to minimize the cost associated to the used resources, maximizing the network load balancing. The optimum placement of VNFs and replicas can provide the optimum locations for the data centers, which will be responsible for the instantiation of VNFs in the network.

\section{Optimization Models}

This section formulates the Link Capacity Dimensioning, TE and RA models as optimization problems subject to a set of constraints, as described next. The notation of all parameters and variables is summarized in Table \ref{parameters}. 

\subsection{Link Capacity Dimensioning Model}

This model allows the initial link dimensioning with the aim to minimize the required capacities for a given topology and a given traffic matrix. Then:
\begin{equation}
\text{Minimize: } \sum_{l \in \vv L} \sum_{t \in \vv T} C_t^l
\end{equation}

The constraint (\ref{linktrafficconstraint}) assures that for the given 
link traffic (left side) the capacity of each link is dimensioned 
according to an over-provisioning factor $\vartheta$. So, $\forall l \in 
\vv L$:

\begin{equation} \label{linktrafficconstraint}
\sum_{\lambda \in \vv \Lambda_b} \sum_{p \in \vv P_{\lambda}} \lambda \cdot R_p \cdot t_p^l
\leq \vartheta \cdot \sum_{t \in \vv T} C_t^l \cdot t
\end{equation}

To be sure that every link is of exactly one bandwidth type:
\begin{equation}
\forall \ell \in \vv L: \sum_t C_t^l = 1
\end{equation}

Finally, the next constraint assures that every traffic demand exactly uses one admissible path:
\begin{equation}
\forall \lambda \in \vv \Lambda_b: \sum_{p \in \vv P_{\lambda}} R_{p}^\lambda = 1
\end{equation}

\subsection{Traffic Engineering Model}

The TE model minimizes the utilization cost of all links in the network given as 

\begin{equation}
Minimize:   \sum_{l \in \vv L} K_l 
\end{equation}

where the cost of every link is related to its link utilization $U_{l}^{TE} $  and defined by the resulting value from all linear cost functions $y_i(U_{l}^{TE}) = a_i \cdot U_{l}^{TE} - b_i$ as follows:

\begin{equation}
\forall l \in \vv L, \forall y \in \vv Y: K_l \geq y \Bigg(  U_{l}^{TE}    \Bigg)
\end{equation}

where constants $a_i$ and $b_i$ are chosen in a way that the slope of the incremental cost values $ y_i \in \vv Y$ approximately follows an exponential function \cite{caria}. The link utilization is given by

\begin{equation}
  U_{l}^{TE} =   \sum_{\lambda \in \vv \Lambda_b} \sum_{p \in \vv P_{\lambda}} \frac{\lambda \cdot R_{p}^\lambda   \cdot t_p^l}{c_l}
\end{equation}

The summation takes into account each traffic demand $\lambda$ out of the set of all background demands $ \vv \Lambda_b$ whose specific path  $p \in \vv  P_{\lambda}$ is traversing link $l$, divided by the link capacity. The only routing constraint for this model assures that every traffic demand exactly uses one admissible path:
\begin{equation*}
\forall \lambda \in \vv \Lambda_b: \sum_{p \in \vv P_{\lambda}} R_{p}^\lambda = 1
\end{equation*}

\begin{table}[!t]
	\renewcommand{\arraystretch}{1.3}
	\caption{Notation}
	\label{parameters}
	\centering
	\footnotesize
	\begin{tabular}{c p{4.2cm}}
		\hline
		\textbf{Parameter} & \textbf{Meaning}\\
		\hline
		$\vv N = \{n_0, n_1, ..., n_{(N-1)}\}$ & set of all nodes\\
		$\vv L = \{l_0, l_1, ..., l_{(L-1)}\}$ & set of all links\\
		$\vv P = \{p_0, p_1, ..., p_{(P-1)}\}$ & set of all paths\\
		$\vv T = \{t_0, t_1, ..., t_{(T-1)}\}$ & set of all bandwidths types\\
		$\vv Y = \{y_0, y_1, ..., y_{(Y-1)}\}$ & set of linear cost functions\\
		$\vv S = \{s_0, s_1, ..., s_{(S-1)}\}$ & set of service chains\\
		$\vv {V_s} = \{v_0, v_1, ..., v_{(V_s-1)}\}$ & set of VNFs in service chain $s$\\
		$\vv {\Lambda} = \{\lambda_0, \lambda_1, ..., \lambda_{(\Lambda-1)}\}$ & set of all traffic demands\\
		$\vv \Lambda_{b} \subseteq \vv \Lambda$ & subset of background traffic demands\\
		$\vv \Lambda_{s} \subseteq \vv \Lambda$ & subset of demands of service chain $s$\\
		$\vv P_{\lambda} \subseteq \vv P$ & subset of paths for a specific $\lambda$\\
		$\vv P_s \subseteq \vv P$ & subset of paths for service chain $s$\\
		$t_p^l \in \{0,1\}$ & 1 if path $p$ traverses link $l$\\
		$r_{v} \in \{0,1\}$ & 1 if function $v$ can be replicated\\
		$r_{max}$ & maximum number of allowed replicas per service chain\\
		$w_{max}$ & maximum number of VNFs per DC\\
		$c_l$ & maximum capacity of link $l$ \\
		$\vartheta$ & over-provisioning capacity ratio\\
		\hline
		\textbf{Variable} & \textbf{Meaning}\\
		\hline
		$K_\ell$ & utilization cost of link $\ell$\\
		$R_p^\lambda \in \{0,1\}$ & 1 if traffic demand $\lambda$ is using path $p$\\
		$C_t^l \in \{0,1\} $ & 1 if link $l$ is of of bandwidth type $t$ \\
		$R_{p}^s \in \{0,1\}$ & 1 if service chain $s$ is using path $p$\\
		$R_{p}^{\lambda,s} \in \{0,1\}$ & 1 if traffic demand $\lambda$ from service chain $s$ is using path $p$\\
		$F_n^{v,s} \in \{0,1\}$ & 1 if VNF $v$ from service chain $s$ is allocated in node $n$\\
		$F_n \in \{0,1\}$ & 1 if node $n$ is running some VNF\\
		\hline
		\vspace{-0.9cm}
	\end{tabular}
\end{table}

\subsection{Resource Allocation Model}
The second optimization model, called RA model, uses a similar objective function than the previous one, but adding a new term using the binary variable $F_n$ ($1$ means the node is used by at min $1$ VNF, $0$ no VNF is assigned) which minimizes the number of nodes that are allocating VNFs subject to the constraints explained below:

\begin{equation}
	Minimize: \alpha \bigg( \sum_{l \in \vv L} K_l \bigg) +  \beta \bigg(\sum_{n \in N} F_n \bigg)
\end{equation}

Both terms are weighted by $\alpha$ and $\beta$ parameters to enable the desired tradeoff between load balancing and network costs. In this paper we restrict to use them as selection parameter for one of the objective functions.

The RA traffic model is different to the TE model and defined as follows. For each service chain  ${s \in \vv S}$ a total  number of $ \| \vv \Lambda_s \|$  traffic demands $\lambda \in \vv \Lambda_s$ are defined, where each of them could not be splitted and have to be forwarded over one service chain specifc path $p \in \vv P_s$, which is taken into account by binary variable $ R_{p}^{\lambda,s}$ (is only $1$ if path $p$ is used by demand $\lambda$). Furthermore, the number of paths $ \| \vv P_s \|$  may be different than the number of demands. Thus, without the usage of VNF replicas all demands of a service chain must use the same path.  As a result, the link utilization due to the RA model is given by:

\begin{equation}
    U_{l}^{RA}   = \sum_{s \in \vv S} \sum_{\lambda \in \vv \Lambda_s} \sum_{p \in \vv P_s} \frac{\lambda \cdot  R_{p}^{\lambda,s}  \cdot t_p^l}{c_l}
\end{equation}

In the RA model, the linear utilization cost functions take into account the superposition of the fix given background TE traffic with the RA traffic  specified as:

\begin{equation}
	\forall l \in \vv L, \forall y \in \vv Y: K_l \geq y \Bigg(     U_{l}^{TE} +   U_{l}^{RA}   \Bigg)
\end{equation} 

Because each demand has to be assigned to a path,  the routing constraint is given by

\begin{equation}
\forall s \in \vv S: \sum_{\lambda \in \vv {\Lambda_s}} \sum_{p \in \vv{P_s}} R_{p}^{\lambda,s} = \| \Lambda_s \|
\end{equation}

Then, to know if an specific node is being used or not, the binary variable $F_n$ will only be $1$  if  $ \geq1$  VNFs are assigned to a node $n$, which is assured by the constraint
\begin{equation}
	\forall n \in \vv N: \frac{\sum_{s \in \vv S} \sum_{v \in \vv V} F_n^{v,s}}{W} \leq F_n \leq  \sum_{s \in \vv S} \sum_{v \in \vv V} F_n^{v,s}
\end{equation}

where the binary variable $ F_n^{v,s}$ indicates, if VNF $v$ from service chain $s$ is allocated to node $n$. $W$ is a large constant to assure, that the left side of the equation is always less than $1$. The next constraint (\ref{routing2}) assures that a certain traffic demand $\lambda$ can only use a path $p$ if the requested service chain is also using the same path. Thus  (\ref{routing2})  takes into account, that more than one demand can use the same path. On the other hand, constraint  (\ref{routing3}) guarantees, that a service schain $s$ can not establish an active path without traffic assigned to it. 
 
\begin{equation} \label{routing2}
\forall s \in \vv S, \forall p \in \vv {P_s},  \forall \lambda \in \vv {\Lambda_s}: R_{p}^{\lambda,s} \leq R_{p}^s
\end{equation}

\begin{equation} \label{routing3}
	\forall s \in \vv S, \forall p \in \vv {P_s}: R_{p}^s \leq \sum_{\lambda \in \vv{\Lambda_s}} R_{p}^{\lambda,s} 
\end{equation}

Furthermore, the number of admissible paths for each service chain $s$ is constrained by the number of replicas. Then, for [$0, 1, 2, ..., r_{max}$] replicas, each service chain can use at maximum [$1,2,3...,(r_{max}+1)$] paths to forward traffic, i.e., 

\begin{equation} \label{rmax}
	\forall s \in \vv S: 1 \leq \sum_{p \in \vv {P_s}} R_{p}^s \leq r_{max} + 1
\end{equation}

Therefore, with no replicas, a certain service chain can only use one path, while increasing number of replicas, the number of admissible paths proportionally increases. Then, for each activated path in service chain $s$ defined by $R_{p}^s$, the next constraint allocates all VNFs of the service chain:
\begin{equation}
 \forall s  \in \vv S,  \forall p \in \vv {P_s},  \forall v \in \vv {V_s}: R_{p}^s \leq \sum_{n \in p} F_n^{v,s}  
\end{equation}

\begin{figure}[!t]
        \centering
	\includegraphics[width=3.0in]{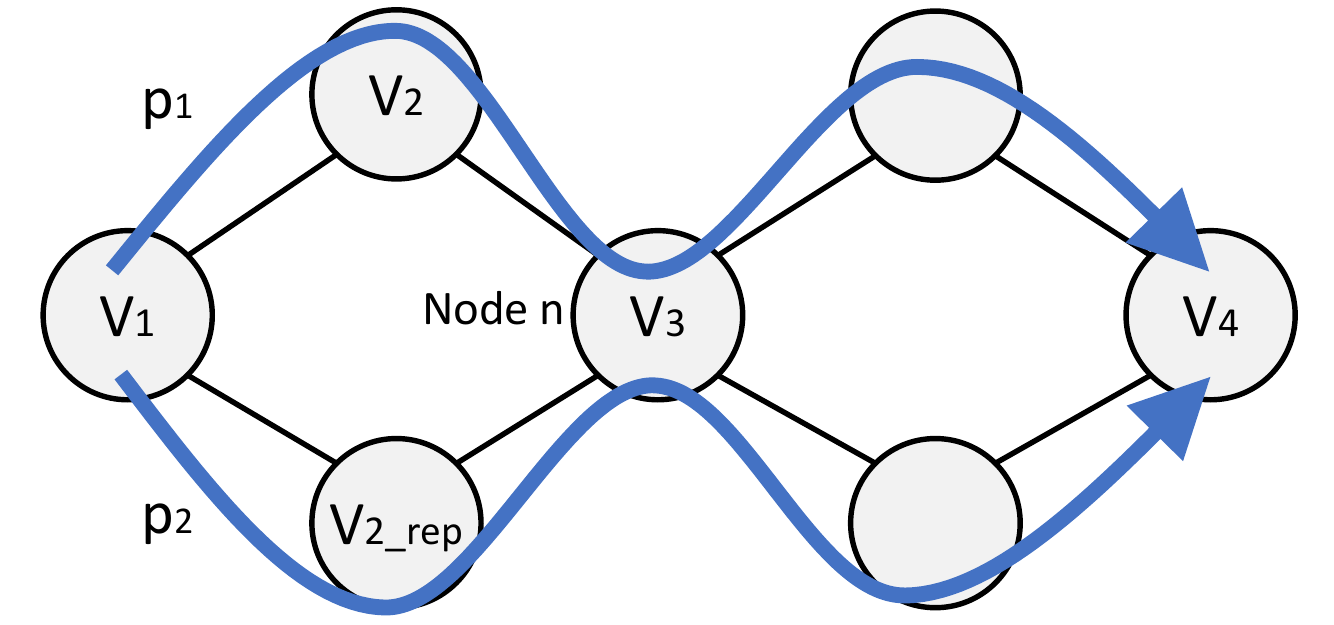}
	\caption{VNF replication strategy}
	\label{strategy}
	\vspace{-0.5cm}
\end{figure}

On the other hand, the sequence order of VNFs in the service chain has to be maintained. Then, for a certain path $p$, the function $v$ can not be allocated in the node $n$, if the previous function $v-1$ is not already allocated in any of the previous nodes of the same path. So, $\forall v \in \vv {V_s}, \forall p \in \vv {P_s}, \forall n \in p$:

\begin{equation}
	\Bigg( \sum_{m = 0}^{n} F_m^{(v-1),s} \Bigg) - F_n^{v,s}  \geq R_{p}^s - 1 \quad when \quad v>0
\end{equation}

The next two constraints limit the maximum number of VNFs that can be allocated in the network. First, the maximum number of VNFs allocated in some specific node $n$ is constrained by the parameter $w_{max}$:
\begin{equation}
	\forall n \in \vv N: \sum_{s \in \vv S} \sum_{v \in \vv {V_s}}  F_n^{v,s} \leq w_{max}
\end{equation}

Second, if a certain function $v$ can be replicated $r_v$, then, the maximum number of replicas is constrained by the maximum number of active paths $R_p^s$. If the function can not be replicated, then can only be placed once. So, $\forall s \in \vv S, \forall v \in \vv {V_s}$:

\begin{equation}
\sum_{n \in N} F_n^{v,s} \leq r_v \sum_{p \in \vv P_s} R_p^s + 1 - r_v
\end{equation}

To improve the load balancing feature using replicated VNFs, we introduce an  additional constraint. In the proposed replication model, we exclude the case where two selected paths $p_1$ and $p_2$ from the service chain $s$ are choosing the same shared node $n$ to place the same function (original and replica). So, following the example shown in Fig. \ref{strategy}, $v_2$ and ${v_2}_{\_rep}$ can never be allocated in the same node $n$. Then, for $\forall v \in \vv {V_s}, \forall p_1 \in \vv {P_s}, \forall p_2 \in \vv {P_s},  \forall n_1 \in p_1, \forall n_2 \in p_2$:
\begin{equation}
R_{p_1}^s + R_{p_2}^s + 2 F_n^{v,s} \cdot r_v \leq 3  \quad  
\left\{ \begin{array}{rl}
&n_1 = n_2 = n \\
&v \neq 0 \neq V_s-1
\end{array} \right.
\end{equation}

To be noted that, this constraint is added for all possible pairs of admissible paths of each service chain. To reduce the computing complexity we restrict, in this work, to link dis-joint paths.

\section{Performance Evaluation}

\begin{figure}[!t]
            \centering
	\includegraphics[width=3.2in]{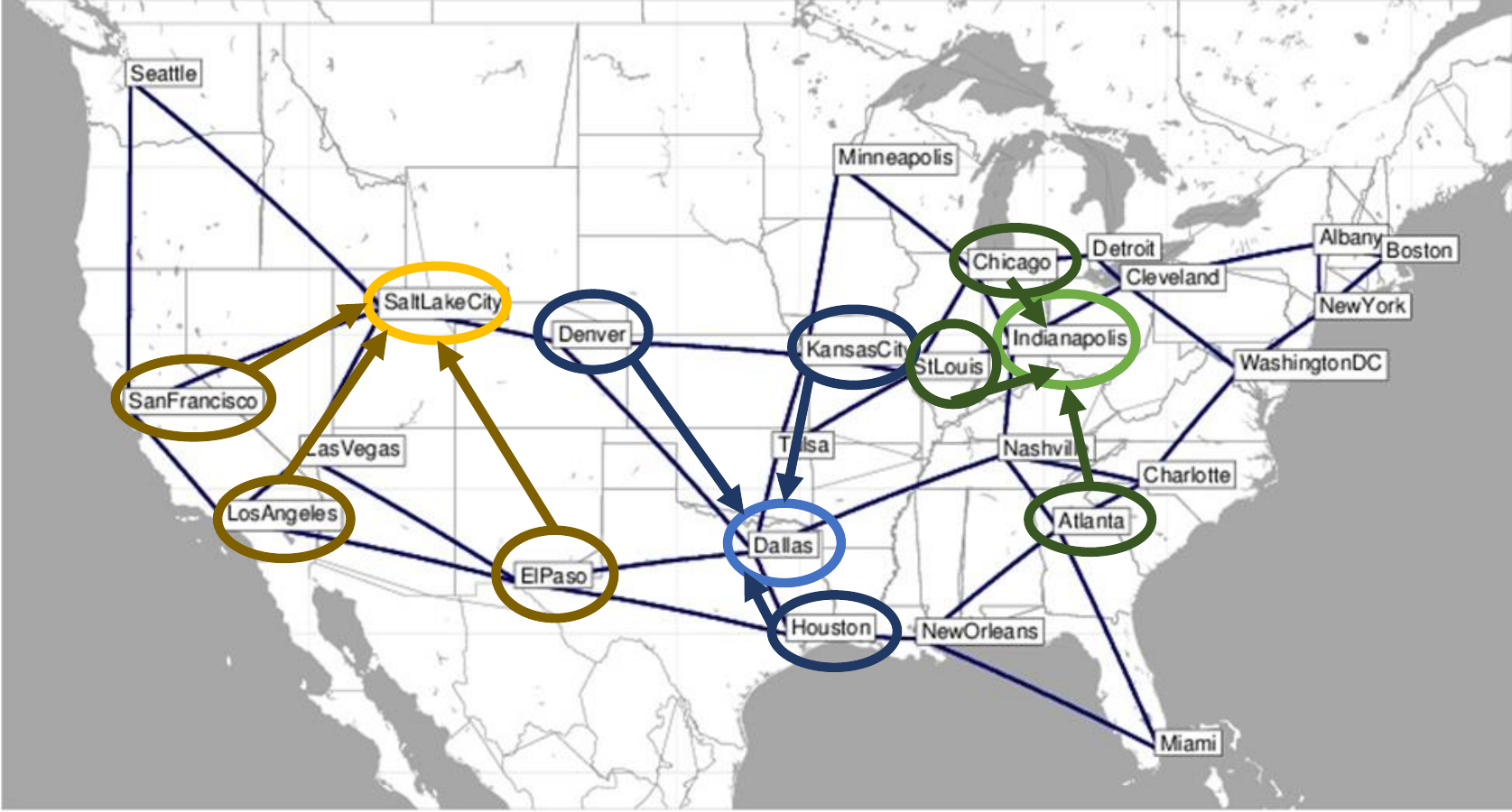}
	\caption{Janos-us network}
	\label{network}
	\vspace{-0.3cm}
\end{figure}

In this section, we show the results for four different optimization scenarios: 1) network load balancing (minLB), 2) minimization of network costs (minNC) given by the number of required data centers (DCs),  3) minimization of network costs with limited number of VNFs per DC (minNC\_constr) and4) network load balancing with limited number of DCs to place VNFs and limited number of VNFs per DC (minLB\_constr). The LP models are implemented using the Gurobi Optimizer \cite{gurobi} and the topology (26 nodes and 84 links)  is chosen from SNDLib website \cite{sndlib} and shown in Fig. \ref{network}.

The background traffic is generated for each source-destination node pair randomly within the interval [1, 4] Gbps. Since the allocation of VNFs is known to be NP-hard, we restrict to 9 S-GWs and 3 P-GWs  placed in the network, which are responsible for the data-center traffic (see Fig.\ref{network}). Unlike other proposals, where the traffic between the S-GW and P-GW is assumed to be non-splittable, each S-NE generates 10 traffic demands of 4.4 Gbps to its associated P-NE, which can be routed along different path. For the link  dimensioning, the background and data-center traffic is optimally load balanced by solving the TE model, where the GWs are not virtualized and placed at the original locations. Based on the traffic flows, we assume an overprovisioning factor of 1.2 to calculate the link capacities $c_l$, which are chosen from different granularities (2.5, 10, 40, 100 and 200 Gbps).

For the RA model the background traffic is routed over the single path derived above, which yields the partial link utilization $U_l^{TE}$. Only the DC traffic is now optimally routed, possibly using parallel paths. All traffic demands of one S/P-NE are handled by one service chain. The service chain analyzed is shown in Fig. \ref{use-case}, which is composed by $VNF_1$ (i.e. S-GW, P-GW, LB), which can not be replicated, $VNF_2$ (i.e. TCP-optimizer) and $VNF_3$ (i.e. Cache PEP), which can be replicated, and finally, $VNF_4$ (i.e. LB, FW, NAT), which again can not be replicated. Depending of the optimization objective, all VNFs may be placed in the same or in different DCs, where also the S/P-NEs are possible locations. Replicated VNFs can be placed only at DCs on different paths. We assume the location of the DCs and the assignment of the VNFs to DCs are the variables to optimize.

\begin{figure*}[!t]
	\centering
        {\includegraphics[width=2.3in]{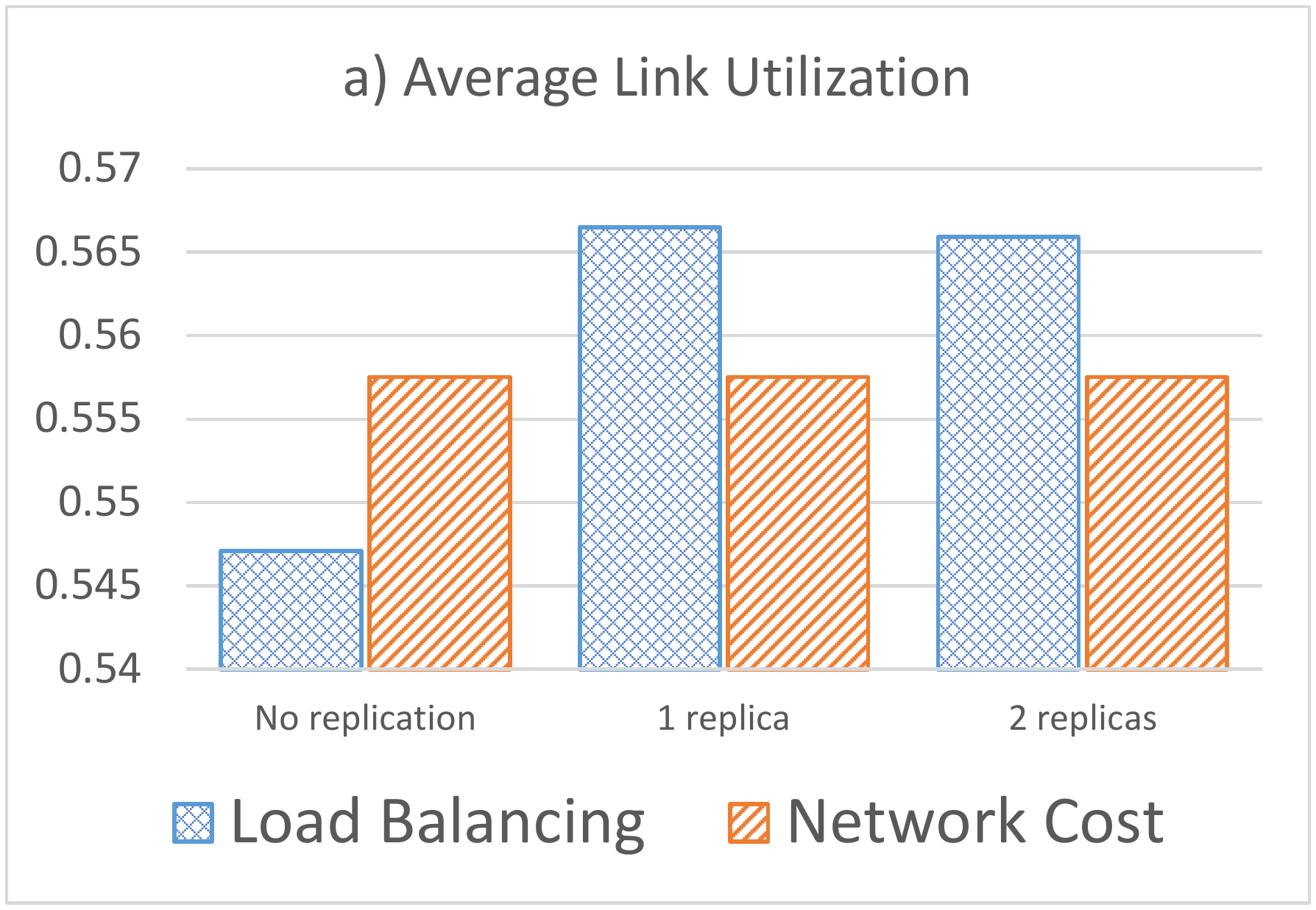}%
		\label{nsf}}
	\hfil
	{\includegraphics[width=2.3in]{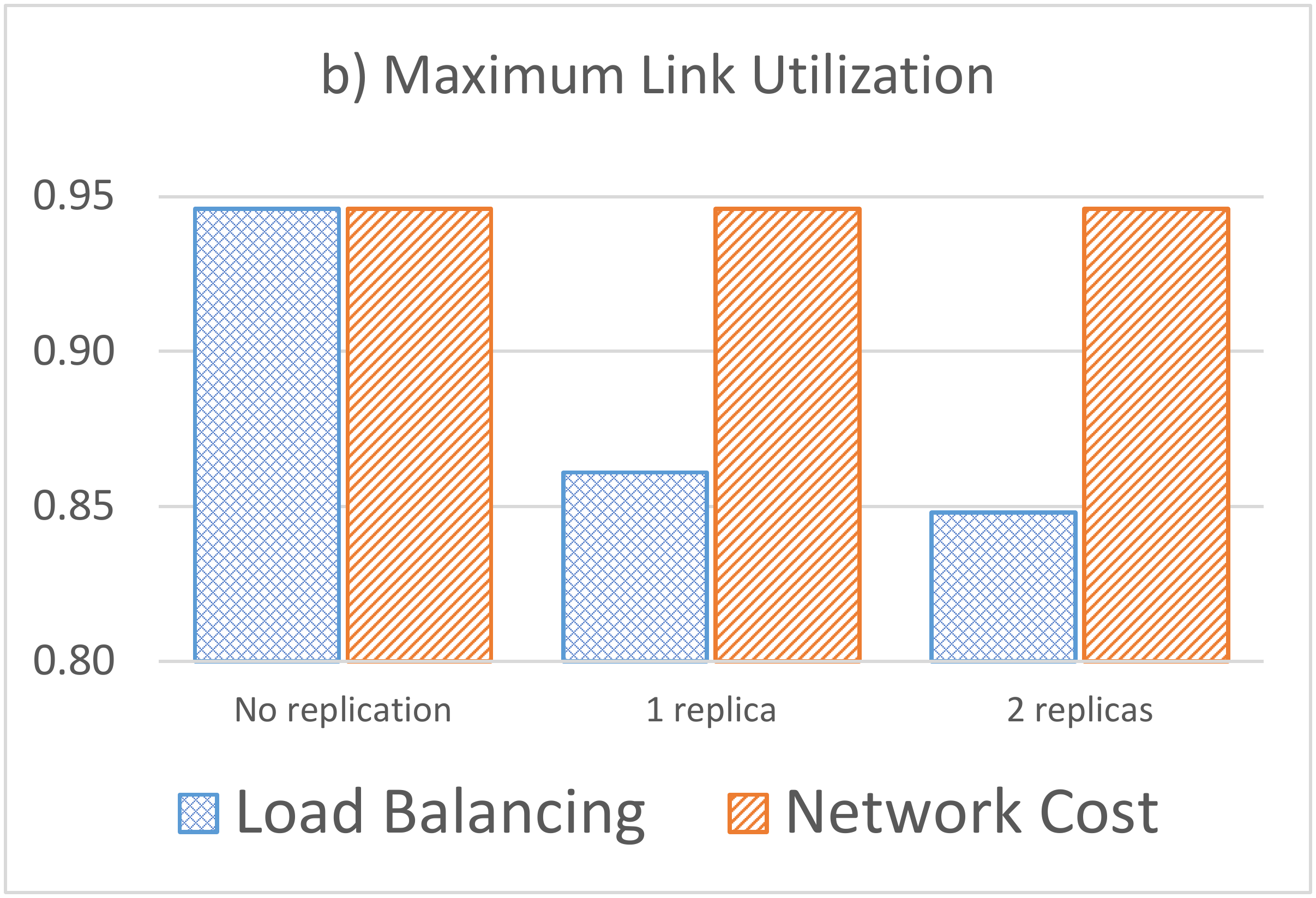}%
		\label{janos}}
	\hfil
	{\includegraphics[width=2.3in]{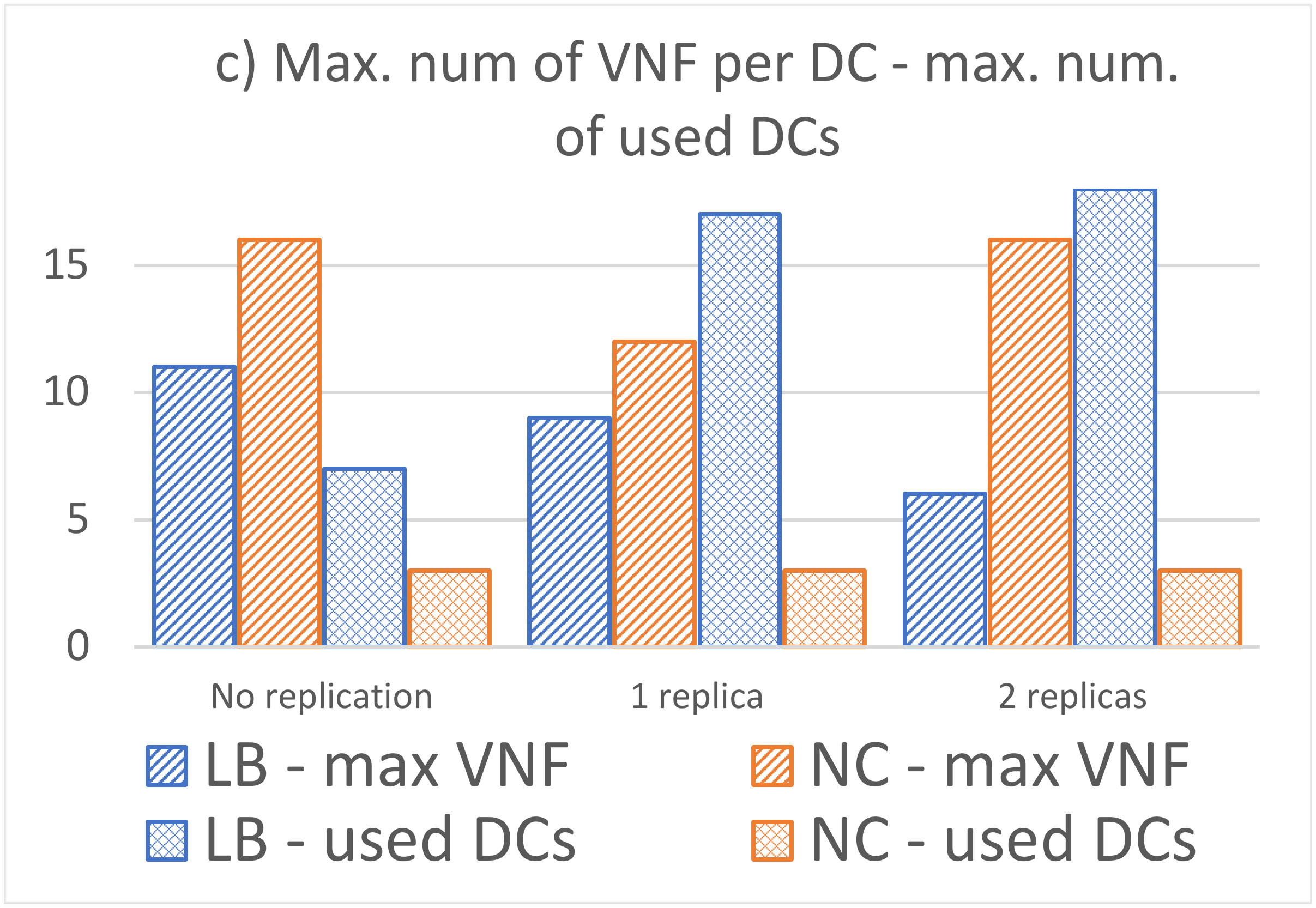}%
		\label{janos}}
	\caption{Comparison between Load Balancing and Network Cost models}
	\label{comparison}
	\vspace{-0.4cm}
\end{figure*}

\subsection{Network Load Balancing vs Minimum Network Node Costs}
We study optimization results obtained for the exclusive minimization of the  load balancing (minLB, $\alpha = 1, \beta = 0$)  and exclusive minimization of the  network costs (minNC, $\alpha = 0, \beta = 1$). In the Fig. \ref{comparison}, we show the comparison of the average link utilization, maximum link utilization, number of used DCs and maximum number of VNFs per DC with no replication, with one and two replicas, respectively. Due to Eq.(\ref{rmax}) the  number of replications are restricted to the maximum value $r_{max}$, later shown in the figures as replicas. 

For the minNC scenario we observe expected results. Due to the cost optimization all VNFs are placed in 3 DCs which are related to the number of P-NEs and only single paths between the S-NEs and P-NEs are used.  Thus the average and maximum link utilization is maintained constant for any case. The maximum number of VNFs per DC for the NC scenario changes if one VNF replication can be used (see Fig. \ref{comparison}c ) although the optimized value of the objective function remains the same. This comes from the fact that the solution is only unique with respect to the number of used DCs and not to the number of assigned VNFs per DC, which is an important behavior and for further studies.

\begin{figure}[!t]
	\includegraphics[width=3.5in]{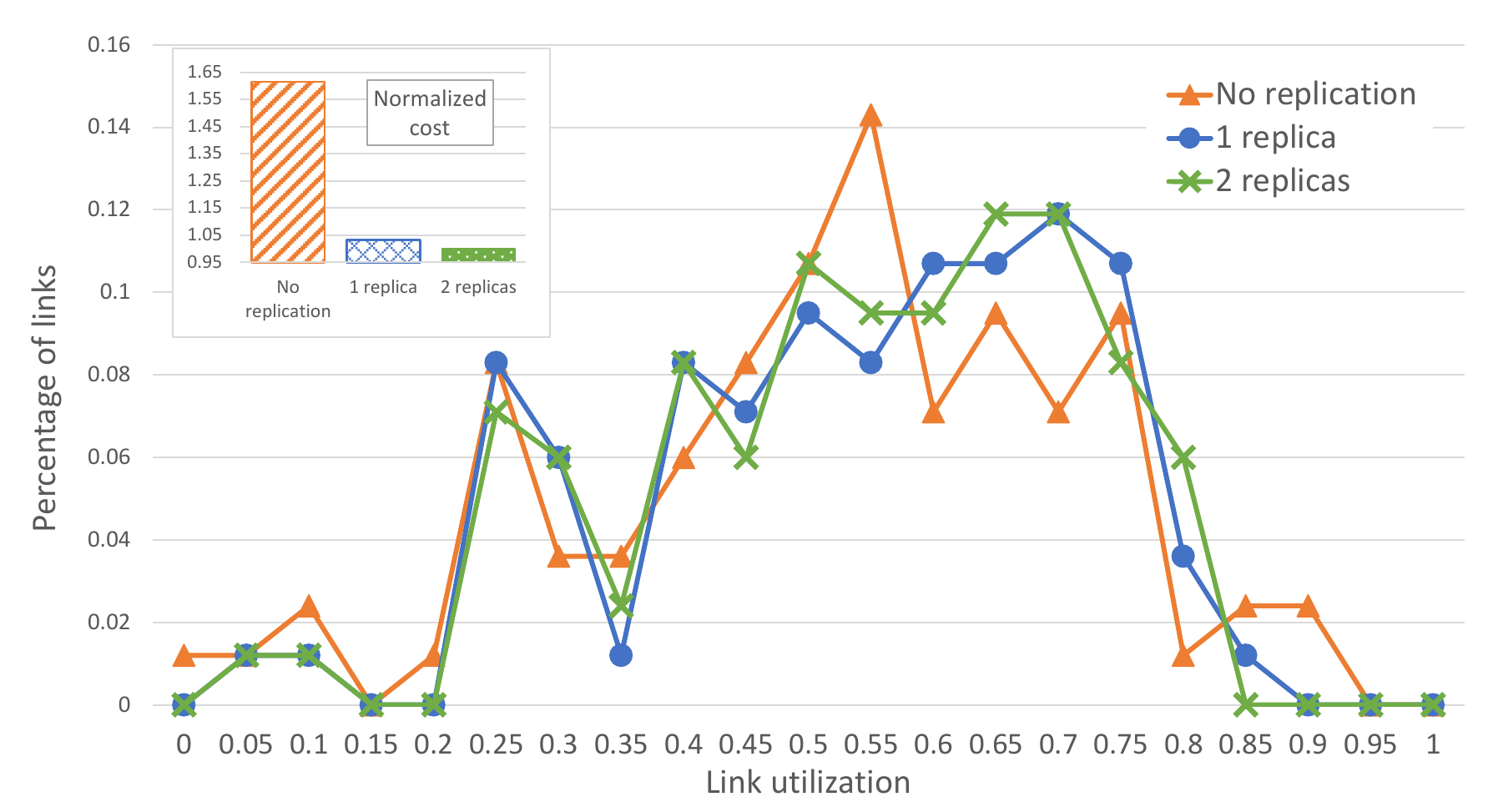}
	\caption{Link Utilization of minLB Model}
	\label{histogram_LB}
	\vspace{-0.4cm}
\end{figure}

While, in the minLB scenario, the average link utilization increases with replication  (Fig. \ref{comparison}a)  because the optimization model tries to decrease number of overloaded links using longer paths with underutilized links, which can be seen from the reduction of the maximum link utilization (Fig. \ref{comparison}b).  In more detail  Fig. \ref{histogram_LB} shows a histogram of the link utilization, which verifies, that an increment of the available parallel paths  to forward traffic decreases the number of overloaded links due to replication. The improvement between one and two replicas is not significant, but it is totally dependent on the chosen topology. The effect on the objective function \ref{objectiveRA} is quite larger, as can be seen from the chart embedded in  Fig. \ref{histogram_LB}.  On the other hand, as shown in  (Fig. \ref{comparison}c), the number of used DCs increases because of longer paths  more DCs are required to allocate replicas. Accordingly the maximum number of VNFs assigned to a DC decreases. 

\begin{table}[!t]
	\renewcommand{\arraystretch}{1.3}
	\vspace{+0.8cm}
	\caption{Node Cost model with at most 8 VNFs per node}
	\label{node_cost}
	\centering
	\footnotesize
	\begin{tabular}{c c c c}
		\hline
		\textbf{Case} & \textbf{avg utilization} & \textbf{path length} &\textbf{avg\_vnf}\\
		\hline
		No replication & 0.568 & 2.22 & 1.38\\
		With replication & 0.55 & 2 & 1.38\\
		\hline
		\vspace{-0.9cm}
	\end{tabular}
\end{table}

\subsection{ Network Optimization under DC and VNF Constraints} 

 In the minNC case, the objective is to minimize the number of used DCs without restricting the number of VNF per DC. However, in a realistic DCs the number of VNFs allocated to a virtual maschine (VM) will be restricted due to performance boundaries of the available computation resources, e.g. the service processing time. Furthermore the number of VMs per DC will be constrained as well.  Due to our small scale example we will deal with this problem by simple constraining the maximum number of VNFs per DC to 8 VNFs. The solution of the constraint optimization problem minNC\_constr results into 6 DCs, which doubles the network costs as shown in Fig.\ref{comparison}. Due to the increased number of DCs the use of VNF replications offers some advantages, as  shown in Table \ref{node_cost}. The constraint reduces the average number of VNFs (avg-vnf) per node and further, the usage of replicas allows to use some alternative paths in parallel, which reduces the average path length (in hops). As a consequence, the total network traffic decreases as well as the average link utilization. 

\begin{figure}[!t]
	\includegraphics[width=3.5in]{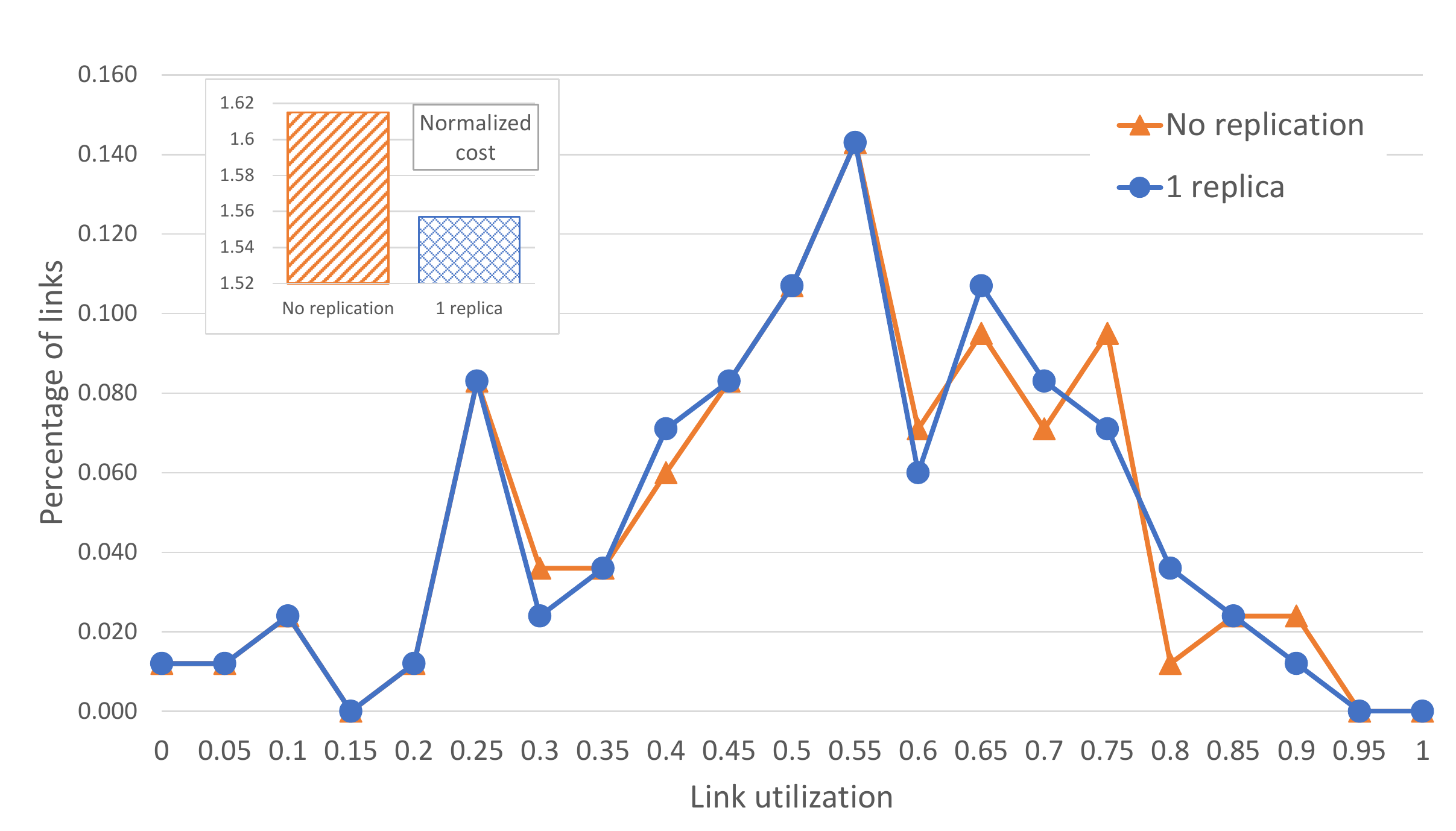}
	\caption{Link Utilization of minLB\_constr  model}
	\label{histogram_LB_constr}
	\vspace{-0.3cm}
\end{figure}

Finally we want to optimize the load balancing under constraints given by the maximum number of usable DCs as well as the maximum number of VNFs per DC, where the constraints are taken from the optimal results given by the minNC\_constr problem.  In the Fig. \ref{histogram_LB_constr}, we show the results for the histogram of the link utilization. Here, we can see that one replica is still able to improve the load balancing, but due to the small example with less effect. 

\section{Conclusions}

As VNFs can only be placed onto servers located in  data centers, the traffic directed to these DCs has not only significant impact on the network costs but also on the load balancing. To enable a cost efficient DC placement by jointly improve the load balancing we introduce the VNF placement with replications in this paper, and especially show how the replications of VNFs can help to load balance the network.

\section*{Acknowledgment}
This work has been performed in the framework of SENDATE-PLANETS (Project ID C2015/3-1), and it is partly funded by the German BMBF (Project ID 16KIS0470).

\bibliographystyle{IEEEtran}
\bibliography{mylib}

\end{document}